\documentclass[prd,showpacs,twocolumn,superscriptaddress,nofootinbib,floatfix,10pt]{revtex4-2}
\usepackage{setspace}
\usepackage[utf8]{inputenc}
\usepackage{float}
\usepackage{amsmath,amssymb,amsfonts,bm}
\usepackage{graphicx}
\usepackage{multirow}
\usepackage[usenames,dvipsnames]{color}
\allowdisplaybreaks
\usepackage[
colorlinks=true,
linkcolor=blue,
breaklinks=true,
urlcolor=blue,
citecolor=blue]{hyperref}
\graphicspath{{figs.dir/}}

\usepackage{orcidlink}

\definecolor{green}{rgb}{0,0.6,0}

\newcommand{\be}{\begin{equation}}
\newcommand{\ee}{\end{equation}}
\newcommand{\bea}{\begin{eqnarray}}
\newcommand{\eea}{\end{eqnarray}}
\newcommand{\beas}{\begin{eqnarray*}}
\newcommand{\eeas}{\end{eqnarray*}}

\newcommand{\veS}{{\bm S}}

\newcommand{\ijs}{\affiliation{Jo{\v z}ef Stefan Institute, Jamova 39, 1000 Ljubljana, Slovenia}}

\newcommand{\lis}{\affiliation{CeFEMA, Center of Physics and Engineering of Advanced Materials, Instituto Superior T{\'e}cnico, Avenida Rovisco Pais 1, 1049-001 Lisboa, Portugal}}

\begin{document}

\title{Extraction of nonperturbative parameters for $D^{(*)}$ mesons from lattice data}

\author{Alexey Nefediev\orcidlink{0000-0002-9988-9430}}\email{Alexey.Nefediev@ijs.si}
\ijs \lis

\begin{abstract}
Recent data for the masses of $D$ and $D^*$ mesons determined using methods of lattice QCD for several values of the charm quark mass different from its physical mass are analysed in heavy quark effective theory. Nonperturbative parameters are extracted that arise at order ${\cal O}(1/m_c)$ in the heavy quark mass expansion of a heavy-light meson mass. The determined parameters are used to establish the charm quark masses corresponding to the employed lattice sets.
\end{abstract}

\maketitle

\section{Introduction}

Hadronic physics of heavy quark flavours is a promising test ground for various studies in nonperturbative QCD. In such studies, heavy-light $D^{(*)}$ and $B^{(*)}$ mesons play an essential role since these are the simplest strongly interacting systems with a well-justified separation of the dynamics related to the heavy and light degrees of freedom. This separation is the essence of
heavy quark effective theory (HQET)---see, for example, book \cite{Manohar:2000dt} and references therein. In particular, since only the hyperfine interaction is relevant for a $S$-wave heavy-light meson then, in the limit $m_Q\to\infty$ (with $Q$ for the heavy quark flavour and $m_Q$ for its mass), the meson mass can be expanded in the inverse powers of $m_Q$ as
\be
m_{hl}=m_Q+\bar{\Lambda}-\frac{1}{2m_Q}(\lambda_1+d_H\lambda_2)+{\cal O}\left(1/m_Q^2\right).
\label{mH}
\ee
Here the coefficient $d_H$ describes the hyperfine splitting,
\be
d_H=\frac32-4(\veS_q\cdot\veS_Q)=3-2S(S+1),
\ee
with $\veS_q$, $\veS_Q$, and $\veS=\veS_q+\veS_Q$ for the spins of the quarks and the total spin, respectively, so
\be
d_H=\left\{
\begin{array}{ll}
+3,&\mbox{for $P$ states with $J^P=0^-$}\\[1mm]
-1,&\mbox{for $P^*$ states with $J^P=1^-$}
\end{array}
\right..
\label{dH}
\ee

The universal parameters $\bar{\Lambda}$, $\lambda_1$, and $\lambda_2$ are $m_Q$ independent and absorb all details of the nonperturbative dynamics in the studied heavy-light meson---they are subject to theoretical or phenomenological determination. The parameter $\bar{\Lambda}$ is associated with the contribution of the light quarks and gluons, $\lambda_1$ is related to the kinetic energy of the heavy quark ($\lambda_1<0$), and the parameter $\lambda_2$ parametrises the heavy quark spin interaction with the chromomagnetic field in the meson.
These nonperturbative parameters appear in the theoretical expressions for the inclusive decay spectra of heavy-light mesons \cite{Shifman:1984wx,Chay:1990da,Bigi:1992su,Bigi:1993fe,Blok:1993va,Manohar:1993qn,Mannel:1993su,Falk:1993dh} and
thus can be extracted directly from experimental data
\cite{Gremm:1996yn,Gremm:1996gg,Falk:1997jq,DELPHI:2005mot}; they can also be related to the Cabibbo-Kobayashi-Maskawa matrix elements \cite{Luke:1993za,Bigi:1993ey,Ball:1994je,Falk:1997gj,Bauer:2000xf}. The most commonly adopted and proven successful methods to determine these nonperturbative parameters of HQET are a theoretical approach based on the QCD sum rules (see, for example, review \cite{Neubert:1993mb} and references therein), lattice calculations (see, for example, Refs.~\cite{Gimenez:1996av,JLQCD:2003chf,Gambino:2017vkx}), and studies of the weak $B$-meson decays
(see, for example, experimental works \cite{CLEO:2001gsa,CLEO:2002woe,CLEO:2001rhd} as well as review \cite{Mannel:1996rg} and book \cite{Manohar:2000dt} and references therein; for a recent study see Ref.~\cite{Mahajan:2023hwe}). The problem of the HQET parameters determination can also be tackled using other nonperturbative theoretical methods such as Dyson--Schwinger equations \cite{Simonov:2000hd,Wang:2008zzx}, quark model approach \cite{Kalashnikova:2001ig}, and so on. It also needs to be mentioned that, unlike $\bar{\Lambda}$ and $\lambda_1$, the parameter $\lambda_2$ depends on the renormalisation scale $\mu$. Hereinafter, charmed mesons are studied, so, if not stated explicitly, $\lambda_2$ is considered at the scale that corresponds to the charm quark mass $m_c$. In this work, we aim at extracting the values of the parameters $\bar{\Lambda}$, $\lambda_1$, and $\lambda_2(m_c)$ from the recent lattice data in Ref.~\cite{Collins:2024sfi}.

\section{Simple estimates of the HQET parameters}

The three intrinsically nonperturbative parameters of HQET admit a straightforward order-of-magnitude estimate,
\bea
&|\lambda_1|\sim\lambda_2\sim\Lambda_{\rm QCD}^2\simeq 0.1~\mbox{GeV}^2,&\nonumber\\[-2mm]
\label{trivial}\\[-2mm]
&\bar{\Lambda}\sim\Lambda_{\rm QCD}\simeq 0.3~\mbox{GeV},&\nonumber
\eea
where here and in what follows the standard nonperturbative scale of QCD is adopted as $\Lambda_{\rm QCD}\simeq 0.3$~GeV.

Another simple but more accurate estimate of the parameter $\lambda_2$ often discussed in the literature is gained by taking the masses of the pseudoscalar and vector mesons following from Eqs.~(\ref{mH}) and (\ref{dH}),
\bea
&&m_P\approx m_Q+\bar{\Lambda}-\frac{\lambda_1+3\lambda_2}{2m_Q},\nonumber\\[-2mm]
\label{ms}\\[-2mm]
&&m_{P^*}\approx m_Q+\bar{\Lambda}-\frac{\lambda_1-\lambda_2}{2m_Q},\nonumber
\eea
and considering the combination
\be
m_{P^*}^2-m_P^2\approx 4\lambda_2\left(1+\frac{\bar{\Lambda}}{m_Q}\right)+{\cal O}\left(\frac{\bar{\Lambda}^2}{m_Q^2}\right).\ee
Therefore, up to the terms suppressed as $\bar{\Lambda}^2/m_Q^2$,
\be
\lambda_2(m_Q)\approx\frac14(m_{P^*}^2-m_P^2)\left(1-\frac{\bar{\Lambda}}{m_Q}\right).
\label{lambda2mQ}
\ee
Then, for the average physical $D^{(*)}$-meson masses \cite{ParticleDataGroup:2022pth},
\bea
m_D^{\rm ph}&=&\frac12(m_{D^0}+m_{D^c})=1.867~\mbox{GeV},\nonumber\\[-2mm]
\\[-2mm]
m_{D^*}^{\rm ph}&=&\frac12(m_{D^{*0}}+m_{D^{*c}})=2.009~\mbox{GeV},\nonumber
\eea
$\bar{\Lambda}\simeq\Lambda_{\rm QCD}$, and the $c$-quark ``running'' mass evaluated in two loops in the $\overline{\rm MS}$ scheme for the renormalisation scale $\mu=m_c$ \cite{ParticleDataGroup:2022pth},
\be
m_c^{\rm ph}=1.27\pm 0.02~\mbox{GeV},
\label{mcPDG}
\ee
one arrives at
\be
\lambda_2(m_c)\simeq 0.1~\mbox{GeV}^2,
\label{simple}
\ee
that agrees well with the order-of-magnitude estimate in Eq.~(\ref{trivial}) above.

It also proves convenient to employ Eqs.~(\ref{mH}) and (\ref{dH}) for the masses of the $D$ and $D^*$ mesons to define their spin-averaged mass and mass splitting as
\begin{align}
&\bar{m}=\frac14(m_D+3m_{D^*})\approx m_c+\bar{\Lambda}-\frac{\lambda_1}{2m_c},\nonumber\\[-2mm]
\label{ms2}\\[-2mm]
&\Delta m=m_{D^*}-m_D\approx \frac{2\lambda_2}{m_c}.\nonumber
\end{align}
The latter relation above allows one to determine
\be
\lambda_2(m_c)\approx\frac12\Delta m^{\rm ph}m_c^{\rm ph}
\ee
that, for $m_c^{\rm ph}$ from Eq.~\eqref{mcPDG} and
\be
\Delta m^{\rm ph}=m_{D^*}^{\rm ph}-m_D^{\rm ph}=0.142~\mbox{GeV},
\ee
gives
\be
\lambda_2(m_c)\approx 0.09~\mbox{GeV}^2,
\label{estimate}
\ee
also in agreement with the previous estimates in Eqs.~\eqref{trivial} and \eqref{simple}.

For $B^{(*)}$ mesons, Eq.~(\ref{lambda2mQ}) gives $\lambda_2(m_b)\simeq 0.11~\mbox{GeV}^2$ that translates to $\lambda_2(m_c)\simeq 0.09~\mbox{GeV}^2$ \cite{Falk:1990pz,Eichten:1990vp}, in good agreement with all estimates above.
It is also relevant to mention that studies of heavy-light mesons performed in the framework of Bethe-Salpeter equations revealed relativistic corrections to relation (\ref{lambda2mQ}) to slightly damp the extracted value of $\lambda_2$ \cite{Wang:2008zzx}.

There is no practical way to reliably estimate the values of the parameters $\lambda_1$ and $\bar{\Lambda}$ separately from the spectrum of hadrons with the charm and bottom quarks. Despite different $m_Q$ scaling of the corresponding terms in the HQET Hamiltonian, in realistic systems, these two parameters enter entangled, and more sophisticated calculations are required to disentangle them.

\begin{table*}[t!]
\caption{Upper panel: masses of the $D$ and $D^*$ mesons for the five lattice sets in Ref.~\cite{Collins:2024sfi}. Middle panel:  spin-averaged masses and splittings defined in Eq.~\eqref{ms2} and evaluated for the lattice masses from the upper panel. Lower panel: the charm quark masses obtained as explained in the text that correspond to the five lattice sets.}
\label{tab:lat}
\begin{ruledtabular}
\begin{tabular}{lccccc}
& Set 1 & Set 2& Set 3& Set 4& Set 5\\
\hline
$m_D^{\rm lat}$, GeV & 1.762(1) & 1.927(1) & 2.064(2) & 2.191(2) & 2.415(2) \\
$m_{D^*}^{\rm lat}$, GeV & 1.898(2) & 2.049(2) & 2.176(2) & 2.294(2) & 2.506(2) \\
\hline
$\bar{m}^{\rm lat}$, GeV & 1.864(2) &2.019(2) &2.148(2) & 2.269(2)& 2.484(2)\\
$\Delta m^{\rm lat} $, MeV & 136(2) & 122(2) & 112(2) & 104(2) & 91(2)\\
\hline
$m_c$, GeV &1.31(3) & 1.46(3) & 1.60(4) & 1.72(4) & 1.96(4)
\end{tabular}
\end{ruledtabular}
\caption{The values of the parameter $\lambda_1$ found in the literature and used in the calculation of its mean value quoted in Eq.~(\ref{lambda1mean}). If several uncertainties of $\lambda_1$ are provided separately in a cited paper, they are summed in quadratures. The value quoted in the last column is extracted in Ref.~\cite{Neubert:1993mb} from the results reported in Ref.~\cite{Eletsky:1991rt}. Since the parameter $\lambda_1$ is scale independent, its values obtained for different systems with heavy quarks can be treated on equal footing. Meanwhile, the value $\lambda_1=-0.5 \pm 0.1$~GeV$^2$ obtained in the sum rules framework in Ref.~\cite{Ball:1993xv} is criticised in Ref.~\cite{Neubert:1993ch}. Also, the result $\lambda_1=-0.24$~GeV$^2$ obtained from semileptonic inclusive $B$-meson decays in Ref.~\cite{Mahajan:2023hwe} does not contain an uncertainty. Thus these two results are not included in the overall fit.}
\label{tab:lambda1}
\begin{ruledtabular}
\begin{tabular}{lccccccccc}
Reference& \cite{Gimenez:1996av} & \cite{CLEO:2001gsa}&\cite{CLEO:2002woe} & \cite{CLEO:2001rhd} & \cite{Kronfeld:2000gk} & \cite{Neubert:1993mb} & \cite{Gremm:1996yn}& \cite{Eletsky:1991rt}&\cite{Chernyak:1996bq}\\
\hline
Method & Lattice & Experiment& Experiment & Experiment & Theory & Theory & Theory& Theory& Theory\\
$-\lambda_1$, GeV$^2$ &0.09(14) &---& 0.25(15) & 0.24(11) & 0.45(12) & 0.25(20)& 0.19(10) & 0.18(6)& 0.14(3)\\
$\bar{\Lambda}$, GeV &---&0.35(13)&0.39(14)&---&$0.68^{+0.02}_{-0.12}$&0.57(7)&0.39(11)&0.50(5)&0.28(4)
\end{tabular}
\end{ruledtabular}
\end{table*}

\section{Extracting nonperturbative parameters from lattice data}

Recently, new lattice data have become available \cite{Collins:2024sfi} for the $DD^*$ phase shifts relevant for the extraction of the pole position of the exotic state $T_{cc}^+$ \cite{LHCb:2021vvq,LHCb:2021auc}.
Compared with the lattice data previously provided in Ref.~\cite{Padmanath:2022cvl}, the new data contain additional points that correspond to the masses of the charm quark different from its physical mass. This made it possible to study the trajectory of the $T_{cc}^+$ pole in the energy complex plane as $m_c$ grows \cite{Collins:2024sfi}. In addition, the new data allow one to investigate the $m_c$ dependence of the $D$ and $D^*$ masses and extract the nonperturbative parameters of HQET in Eq.~\eqref{mH}.

It is important to notice that the lattice data in Ref.~\cite{Collins:2024sfi} correspond to an unphysically large pion mass $m_\pi^{\rm lat}\approx 0.28$~GeV. Then, in order to proceed to the physical point in $m_\pi$, one is to employ chiral extrapolation and apply to the lattice $D^{(*)}$-meson masses a correction factor derived in chiral perturbation theory \cite{Cleven:2010aw},
\be
\eta_{\rm extr}=1+h_1\left(\frac{m_\pi^{\rm ph}}{m_{D^{(*)}}^{\rm ph}}\right)^2\left[\left(\frac{m_\pi^{\rm lat}}{m_\pi^{\rm ph}}\right)^2-1\right]\approx 1.007,
\label{eta}
\ee
where $h_1=0.42$ while $m_\pi^{\rm ph}$ and $m_{D^{(*)}}^{\rm ph}$ are the physical masses of the pion and $D^{(*)}$ meson, respectively. The deviation of this factor from unity lies within 1\%, so it is included into the overall uncertainty of the results.

It has to be noticed that the evolution of the $m_D^{\rm lat}$ and $m_{D^*}^{\rm lat}$ in Ref.~\cite{Collins:2024sfi} is provided in terms of the hopping parameter $\kappa_c$---an auxiliary lattice parameter
related to the bare charm quark mass---which makes investigations of $m_c$ dependence of physical parameters not straightforward. Nevertheless, one can proceed by simultaneously considering the $D$- and $D^*$-meson masses from a given lattice set in Table~\ref{tab:lat} that correspond to the same (though yet unknown) charm quark mass. This way, a lattice dependence $\bar{m}^{\rm lat}(\Delta m^{\rm lat})$ can be established that consists of five points. These points can be fitted with the theoretical curve
\be
\bar{m}(\Delta m)=\bar{\Lambda}-\frac{\lambda_1}{4\lambda_2}\Delta m +\frac{2\lambda_2}{\Delta m}
\label{fitfunc}
\ee
derived from the two relations in Eq.~\eqref{ms2}.

Let us start from a qualitative analysis of relation \eqref{fitfunc}. Using the estimates from Eq.~(\ref{trivial}), one can observe that the three terms on the right-hand side of Eq.~(\ref{fitfunc}) are expected to take utterly different values: the major contribution is provided by the last term while the first and second terms contribute about 15\% and 1\%, respectively. It implies that one cannot expect to accurately extract the parameter $\lambda_1$ using relation (\ref{fitfunc}). Meanwhile, once the $\lambda_2$- and $\bar{\Lambda}$-dependent terms in Eq.~(\ref{fitfunc}) provide the dominant and sizeable subdominant contributions to the dependence $\bar{m}(\Delta m)$, respectively, it should be possible to reliably extract the parameter $\lambda_2$ and obtain a fairly accurate estimate for $\bar{\Lambda}$.

\begin{figure}[t!]
\centering
\includegraphics[width=0.45\textwidth]{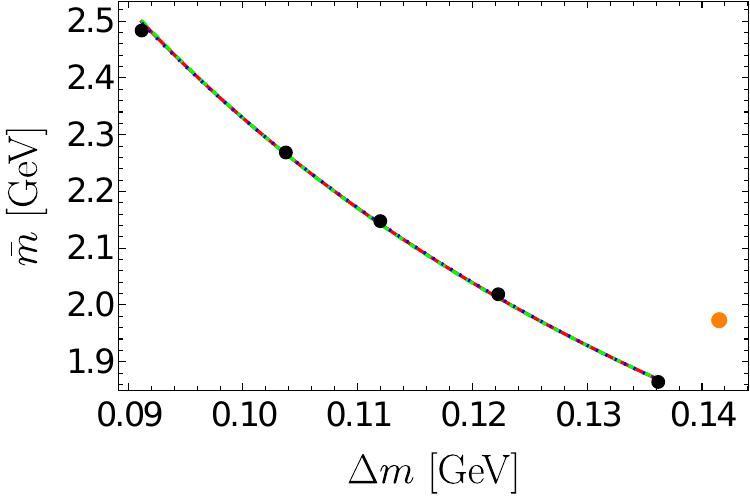}
\caption{The theoretical dependence $\bar{m}(\Delta m)$ from Eq.~(\ref{fitfunc}) fitted to
to the lattice points (black filled dots) from Table~\ref{tab:lat}. The (indistinguishable by eye) blue, red, and green dashed lines are for fits~I, II, and III, respectively. The big orange dot corresponds to the spin-averaged mass evaluated for the physical $D$ and $D^*$ masses~\cite{ParticleDataGroup:2022pth}.}
\label{fig:fit}
\end{figure}

If all three quantities, $\{\bar{\Lambda},\lambda_1,\lambda_2\}$, are simultaneously treated as fitting parameters, the fit to the five lattice points returns an unphysical value of $\lambda_1>0$ (that implies a negative kinetic energy of the heavy quark) while the fit constrained by the condition $\lambda_1<0$ naturally hits the boundary and returns $\lambda_1=0$. Meanwhile, as explained above, although we anticipate a strong correlation between the parameters $\bar{\Lambda}$ and $\lambda_1$, the parameter $\lambda_2$ is expected to be only weakly dependent on $\lambda_1$.
Then, in order to proceed, we consider a representative set for the values of the parameter $\lambda_1$ found in the literature and obtained employing different methods---see Table~\ref{tab:lambda1}. Averaging these results (that consists in minimising $\chi^2(\lambda_1)=\sum_n(\lambda_1-\lambda_1^{(n)})^2/(\delta\lambda_1^{(n)})^2$, with $n$ enumerating the values quoted in Table~\ref{tab:lambda1}), we find
\be
\bar{\lambda}_1=-0.17\pm 0.03~\mbox{GeV}^2,
\label{lambda1mean}
\ee
that we treat as the central value $\lambda_1^{\rm cent}$ and allow $\lambda_1$ to deviate sufficiently strongly from it, with $\delta\lambda_1=\lambda_1^{\rm cent}-\lambda_1^{\rm min}=\bar{\lambda}_1$, with $\lambda_1^{\rm min}=0$ consistent with the result of the three-parameter fit above. Therefore, in fits~I, II, and III, we preset $\lambda_1=0$, $\lambda_1=\bar{\lambda}_1$, and $\lambda_1=2\bar{\lambda}_1$, respectively. After that, $\bar{\Lambda}$ and $\lambda_2$ are fitted to the lattice data. The results are quoted in Table~\ref{tab:fitsIandII} and shown in Fig.~\ref{fig:fit}. One can conclude that all three fits describe the lattice data equally well and return similar values for the parameter $\lambda_2$, as quoted in Table~\ref{tab:fitsIandII}. Then, averaging the values obtained for the latter in fits~I, II, and III, we arrive at the final result,
\be
\lambda_2(m_c)=0.090\pm 0.002~\mbox{GeV}^2,
\label{lambda2final}
\ee
where the uncertainty is obtained as a sum in quadratures of
$\delta\lambda_2^{\rm fit}=1.8\cdot 10^{-3}$~GeV$^2$ coming from averaging over the fitted values of $\lambda_2$ in Table~\ref{tab:fitsIandII} including correlations, and $\delta\lambda_2^{\rm extr}=\lambda_2(\eta_{\rm extr}-1)\approx 0.6\cdot 10^{-3}$~GeV$^2$ for the chiral extrapolation, with $\eta_{\rm extr}$ introduced in Eq.~\eqref{eta}.

The result \eqref{lambda2final} complies well with the simple estimates in Eqs.~\eqref{trivial}, \eqref{simple}, and \eqref{estimate} above and supersedes them. It is also compatible with the value $\lambda_2(2\bar{\Lambda})=0.12\pm 0.02$~GeV$^2$ obtained in the QCD sum rules technique and quoted in the review \cite{Neubert:1993mb}. Since the scale-dependent parameter $\lambda_2$ is evaluated here for $\mu\simeq m_c$, its direct comparison with the results obtained in the literature from $B$ mesons using the methods of lattice QCD \cite{Gimenez:1996av,JLQCD:2003chf,JLQCD:2003chf} is not straightforward.

\begin{table}[t!]
\caption{The values of the HQET parameters for fits I-VI. As explained in the text, either $\lambda_1$ (fits I-III) or $\lambda_2$ (fits IV-VI) is preset to take the value marked with asterisk and quoted in the second or third column, respectively.}
\label{tab:fitsIandII}
\begin{ruledtabular}
\begin{tabular}{lccc}
Fit & $\lambda_1$, GeV$^2$ &$\lambda_2$, GeV$^2$& $\bar{\Lambda}$, GeV\\
\hline
Fit I   & $0^*$     & 0.087(2) & 0.595(25)\\
Fit II  & $-0.17^*$ & 0.090(2) & 0.485(27)\\
Fit III & $-0.34^*$ & 0.093(2) & 0.382(28)\\
\hline
Fit IV  & -0.05(8)  &$0.088^*$ & 0.556(28)\\
Fit V   & -0.16(9)  &$0.090^*$ & 0.487(31)\\
Fit VI  & -0.27(10) &$0.092^*$ & 0.417(34)
\end{tabular}
\end{ruledtabular}
\end{table}

As anticipated, the parameter $\bar{\Lambda}$ is much more sensitive to the value of $\lambda_1$. Its variation between fits I and II as well as between fits II and III in Table~\ref{tab:fitsIandII}, that constitutes about 25\%, complies well with a simple estimate following from the first relation in Eq.~(\ref{ms2}). Indeed, in the combination $\bar{\Lambda}-\lambda_1/(2m_c)$, a change in $\lambda_1$ at the level of 0.2~GeV$^2$ can be recast in the shift in $\bar{\Lambda}$ of the order 0.1~GeV. With the value of $\lambda_2$ fixed as given in Eq.~\eqref{lambda2final}, we can perform extra fits to the lattice sets to disentangle the parameters $\bar{\Lambda}$ and $\lambda_1$. The results obtained for the three preset values of $\lambda_2$ consistent with Eq.~\eqref{lambda2final} are quoted in Table~\ref{tab:fitsIandII} (fits~IV-VI). Treating the spreads in the central values of $\bar{\Lambda}$ and $\lambda_1$ between fits~IV-VI as model errors and summing them in quadratures with the corresponding statistical uncertainties quoted in parentheses in Table~\ref{tab:fitsIandII}, we find
\begin{align}
\bar{\Lambda}&=0.49\pm 0.08~\mbox{GeV},\nonumber\\[-2mm]
\label{Ll1final}\\[-2mm]
\lambda_1&=-0.16\pm 0.15~\mbox{GeV}^2.\nonumber
\end{align}
The central value of $\lambda_1$ above agrees with the central value of $\bar{\lambda}_1$ in Eq.~\eqref{lambda1mean}, and the uncertainty is compatible with the deviations of $\lambda_1$ from $\bar{\lambda}_1$ in fits I and III in Table~\ref{tab:fitsIandII}. It can be regarded as a self-consistency check of the approach. Then, the obtained value of $\bar{\Lambda}$ in Eq.~\eqref{Ll1final} complies well with the values found in the literature and quoted in Table~\ref{tab:lambda1}. Its
uncertainty $\simeq 16\%$ should not come as a surprise given the discussion after Eq.~\eqref{fitfunc}. Therefore, additional data are needed to more reliably disentangle the parameters $\bar{\Lambda}$ and $\lambda_1$.

With the fitted value of $\lambda_2$ provided in Eq.~(\ref{lambda2final}), we are in a position to employ the second relation in Eq.~(\ref{ms2}) (that is free of the problem with the entangled parameters $\bar{\Lambda}$ and $\lambda_1$) and evaluate the charm quark masses that correspond to the five lattice sets provided in Table~\ref{tab:lat}; the result is presented in the lower row of the same table.

We notice that the lattice data in Ref.~\cite{Collins:2024sfi} demonstrate a tension between the values of the spin-averaged mass and splitting in Table~\ref{tab:lat}. It is evidenced by Fig.~\ref{fig:fit} where the big orange dot, representing the physical $D^{(*)}$-meson masses, hits neither the fitting curves nor their continuation. The situation is likely to improve in future lattice updates, if the lattice action is taken to higher orders in the lattice spacing \cite{private}. It can also potentially facilitate more accurate determination of the parameters $\bar{\Lambda}$ and $\lambda_1$ from the lattice data. Meanwhile, it would be natural to expect the corresponding corrections to the $D$ and $D^*$ masses to be similar in magnitude and weakly $m_c$ dependent (applying such corrections would result in the overall shift of the black points and fitting curves in Fig.~\ref{fig:fit} by $\delta\bar{m}\approx 0.15$~GeV upward). Then, they should cancel to a large extent in the mass splitting $\Delta m$ and leave intact the determined parameter $\lambda_2$ in Eq.~(\ref{lambda2final}) and the extracted charmed quark masses in Table~\ref{tab:lat}.

\section{Conclusions}

In this short paper, we analysed the recent lattice data on the $D^{(*)}$-meson masses obtained for several unphysical values of the charm quark mass using the methods of lattice QCD. We extracted the nonperturbative parameters that appear in the HQET expansion of a heavy-light meson mass to quantify the contribution of the light quarks and gluons, the kinetic energy of the heavy quark, and the heavy quark spin interaction with the chromomagnetic field in the meson. The extracted values of the nonperturbative parameters comply well with both simple quantitative estimates and other results found in the literature. However, two of them, $\bar{\Lambda}$ and $\lambda_1$, are strongly entangled and as a result are determined with large uncertainties. The situation is likely to improve with further lattice updates. The parameter $\lambda_2$ is determined more accurately. It is most difficult for theoretical determination, so establishing its value is an important task for nonperturbative QCD. Employing the extracted value of $\lambda_2$, the charmed quark masses corresponding to the five lattice sets are determined. They can be employed in various quark model calculations.

\acknowledgments

The authors is grateful to Sa{\v s}a Prelovsek for clarifying discussions on lattice techniques and thanks the Institite of Theoretical Physics of Chinese Academy of Sciences, where this work was finalised, for warm hospitality.
This work is supported by the Slovenian
Research Agency (research core Funding No. P1-0035) and by CAS President’s International Fellowship Initiative (PIFI) (Grant No. 2024PVA0004).


\end{document}